\documentclass[final]{agujournal2019}
\usepackage{url} %this package should fix any errors with URLs in refs.
\usepackage{lineno}
\usepackage[inline]{trackchanges} %for better track changes. finalnew option will compile document with changes incorporated.
\usepackage{soul}
\usepackage{amssymb}
\usepackage{amsmath}
\nolinenumbers
\draftfalse

\journalname{Journal of Advances in Modeling Earth Systems (JAMES)}

\begin{document}

%% ------------------------------------------------------------------------ %%
%  Title
%
% (A title should be specific, informative, and brief. Use
% abbreviations only if they are defined in the abstract. Titles that
% start with general keywords then specific terms are optimized in
% searches)
%
%% ------------------------------------------------------------------------ %%

% Example: \title{This is a test title}

\title{On the choice of training data for machine learning of geostrophic mesoscale turbulence}

%% ------------------------------------------------------------------------ %%
%
%  AUTHORS AND AFFILIATIONS
%
%% ------------------------------------------------------------------------ %%

% Authors are individuals who have significantly contributed to the
% research and preparation of the article. Group authors are allowed, if
% each author in the group is separately identified in an appendix.)

% List authors by first name or initial followed by last name and
% separated by commas. Use \affil{} to number affiliations, and
% \thanks{} for author notes.
% Additional author notes should be indicated with \thanks{} (for
% example, for current addresses).

% Example: \authors{A. B. Author\affil{1}\thanks{Current address, Antartica}, B. C. Author\affil{2,3}, and D. E.
% Author\affil{3,4}\thanks{Also funded by Monsanto.}}

\authors{F. E. Yan\affil{1}, J. Mak\affil{1,2}, Y. Wang\affil{1,2}}

% \affiliation{1}{First Affiliation}
% \affiliation{2}{Second Affiliation}
% \affiliation{3}{Third Affiliation}
% \affiliation{4}{Fourth Affiliation}

\affiliation{1}{Department of Ocean Science, Hong Kong University of Science and Technology}
\affiliation{2}{Center for Ocean Research in Hong Kong and Macau, Hong Kong University of Science and Technology}
%(repeat as many times as is necessary)

%% Corresponding Author:
% Corresponding author mailing address and e-mail address:

% (include name and email addresses of the corresponding author.  More
% than one corresponding author is allowed in this LaTeX file and for
% publication; but only one corresponding author is allowed in our
% editorial system.)

% Example: \correspondingauthor{First and Last Name}{email@address.edu}

\correspondingauthor{Fei Er Yan}{feyan@connect.ust.hk}
\correspondingauthor{Julian Mak}{julian.c.l.mak@googlemail.com}

%% Keypoints, final entry on title page.

%  List up to three key points (at least one is required)
%  Key Points summarize the main points and conclusions of the article
%  Each must be 140 characters or fewer with no special characters or punctuation and must be complete sentences

% Example:
% \begin{keypoints}
% \item	List up to three key points (at least one is required)
% \item	Key Points summarize the main points and conclusions of the article
% \item	Each must be 140 characters or fewer with no special characters or punctuation and must be complete sentences
% \end{keypoints}

\begin{keypoints}
\item Investigate dependence of convolution neural network's on choice of training data for geostrophic turbulence
\item Eddy force function used as a way to filter out dynamically inert eddy fluxes
\item Models trained on filtered eddy fluxes at least as accurate but more robust than models trained on the divergence of eddy fluxes

%\begin{verbatim}
%          ================================================
%                |\      _,,,---,,_
%          ZZZzz /,`.-'`'    -.  ;-;;,_
%               |,4-  ) )-,_. ,\ (  `'-'
%              '---''(_/--'  `-'\_)         Here, Miffy was
%          ================================================
%\end{verbatim}
\end{keypoints}

%% ------------------------------------------------------------------------ %%
%
%  ABSTRACT and PLAIN LANGUAGE SUMMARY
%
% A good Abstract will begin with a short description of the problem
% being addressed, briefly describe the new data or analyses, then
% briefly states the main conclusion(s) and how they are supported and
% uncertainties.

% The Plain Language Summary should be written for a broad audience,
% including journalists and the science-interested public, that will not have 
% a background in your field.
%
% A Plain Language Summary is required in GRL, JGR: Planets, JGR: Biogeosciences,
% JGR: Oceans, G-Cubed, Reviews of Geophysics, and JAMES.
% see http://sharingscience.agu.org/creating-plain-language-summary/)
%
%% ------------------------------------------------------------------------ %%

%% \begin{abstract} starts the second page

\begin{abstract}
`Data' plays a central role in data-driven methods, but is not often the subject
of focus in investigations of machine learning algorithms as applied to Earth
System Modeling related problems. Here we consider the case of eddy-mean
interaction in rotating stratified turbulence in the presence of lateral
boundaries, a problem of relevance to ocean modeling, where the eddy fluxes
contain dynamically inert rotational components that are expected to contaminate
the learning process. An often utilized choice in the literature is to learn
from the divergence of the eddy fluxes. Here we provide theoretical arguments
and numerical evidence that learning from the eddy fluxes with the rotational
component appropriately filtered out results in models with comparable or better
skill, but substantially improved robustness. If we simply want a data-driven
model to have predictive skill then the choice of data choice and/or quality may
not be critical, but we argue it is highly desirable and perhaps even necessary
if we want to leverage data-driven methods to aid in discovering unknown or
hidden physical processes within the data itself.
\end{abstract}

\section*{Plain Language Summary}

Data-drive methods and machine learning are increasingly being utilized in
various problems relating to Earth System Modeling. While there are many works
focusing on the machine learning algorithms or the problems themselves, there
has been relative few investigations into the impact of data choice or quality,
given the central role the data plays. We consider here the impact of data
choice for a particular problem of eddy-mean interaction of relevance to ocean
modeling, and provide theoretical arguments and numerical evidence to suggest
that one choice (informed by our theoretical understanding of the underlying
problem) is preferable over a more standard choice utilized in the literature.
While the choice of data choice and/or quality may not be critical if we simply
want a data-driven model to `work', we argue it is highly desirable (possibly
even a necessity) if we want to go beyond having models that just `work', such
as leveraging data-driven methods to help us in discovering unknown or hidden
physical processes within the data itself.

%% ------------------------------------------------------------------------ %%
%
%  TEXT
%
%% ------------------------------------------------------------------------ %%

%%% Suggested section heads:
% \section{Introduction}
%
% The main text should start with an introduction. Except for short
% manuscripts (such as comments and replies), the text should be divided
% into sections, each with its own heading.

% Headings should be sentence fragments and do not begin with a
% lowercase letter or number. Examples of good headings are:

% \section{Materials and Methods}
% Here is text on Materials and Methods.
%
% \subsection{A descriptive heading about methods}
% More about Methods.
%
% \section{Data} (Or section title might be a descriptive heading about data)
%
% \section{Results} (Or section title might be a descriptive heading about the
% results)
%
% \section{Conclusions}

%Text here ===>>>

%\cite<e.g.>[post note]{Marshall-et-al12}
%\citeA{Rathgeber-et-al17}

%%%%%%%%%%%%%%%%%%%%%%%%%%%%%%%%%%%%%%%%%%%%%%%%%%%%%%%%%%%%%%%%%%%%%%%%%%%%%%%%

%-------------------------------------------------------------------------------

\section{Introduction}\label{sec:intro}

Data-driven methods and machine learning algorithms are increasingly being
utilized in problems relating to Earth system and/or climate modeling, and there
is no doubt such methods have a strong potential in greatly enhancing model
skill and/or reducing computation cost in various numerical models. Some
examples of usage includes dynamical processes in the atmosphere
\cite<e.g.,>{BrenowitzBretherton19, YuvalOGorman20, Mooers-et-al21,
Connolly-et-al23, Sun-et-al23}, climate modeling \cite<e.g.,>{Bescombes-et-al21,
SonnewaldLguensat21}, see ice prediction \cite<e.g.,>{Bolibar-et-al20,
Andersson-et-al21}, identification problems in oceanography
\cite<e.g.,>{Jones-et-al19, Thomas-et-al21, Sonnewald-et-al19,
Sonnewald-et-al23}, and our primary focus here, on ocean mesoscale turbulence
\cite<e.g.,>{BoltonZanna19, ZannaBolton20, GuillauminZanna21}. We refer the
reader to the works of \citeA{Reichstein-et-al19}, \citeA{Irrgang-et-al21},
\citeA{Sonnewald-et-al21} and \citeA{CampsValls-et-al23} for a more
comprehensive review.

One criticism of some data-driven methods and machine learning algorithms is the
`black-box' nature of the resulting models. In general, for a problem with input
$x$ and output $y$, a focus of data-driven methods is to find some mapping $f$
such that $f(x) = y$, where $f$ could be deterministic or probabilistic
depending on the algorithm used to obtain $f$. The lack of interpretability for
$f$ in certain instances brings into question several important issues with the
use of data-driven methods. The first is robustness and applicability in
different regimes: are the models doing the right things for the `right' reasons
(or at least not the `wrong' ones)? If for the `wrong' reasons, then it is
perfectly plausible that trained up models can behave erratically when taken
outside the trained regimes and, given the nonlinear and convoluted nature of
the model itself, generate subtly wrong results that might be close to
impossible to check. The second relates to further utilities of the methods
themselves: is it possible to use such methods to aid process discovery from the
data itself? A lack of interpretability would suggest a negative answer to that
question. With that in mind, there has been an increasing focus on physically
constrained and/or interpretable/explainable models \cite<e.g.,>{ZhangLin18,
Brenowitz-et-al20, ZannaBolton20, Beucler-et-al21, Kashinath-et-al21,
SonnewaldLguensat21, Yuval-et-al21, Barnes-et-al22, Clare-et-al22b,
LopezGomez-et-al22, Guan-et-al23}. While the tools and algorithms do exist, this
is a fundamentally harder problem, since the training step ultimate becomes one
of a constrained optimization problem.

While the algorithms and nature of the resulting model $f$ (e.g. linear vs.
nonlinear, generative vs. discriminative, model complexity) are important
details, at the very base level we are really dealing with the problem of
\emph{data regression}. We would thus expect \emph{data choice} and/or
\emph{data quality} to critically affect the training, the performance or the
useful information that could be extracted/encoded by the model, but are issues
that have not received much investigation. If we simply want a model that
`works' in the sense of producing a `skilled' prediction in whatever metric we
think is relevant, then the issue of data quality and/or content may not be
critical, since we are simply looking for some optimal fit. If on the other hand
we are interested in the harder problem of optimal fit with constraints, such as
having a model that works for the `right' reasons (e.g. satisfying physical
conservation laws), or using data-driven methods for process discovery (e.g.
telling us about the underlying physics of a problem), then one might imagine
the choice and quality of data exposed to the model should be important.
Furthermore, certain data may be more accessible for the machine learning
algorithms to extract/predict features from (e.g. smoothness and/or
spatio-temporal scale of data), which has practical consequences for the
optimization procedure at the model training and prediction step.

To demonstrate that not all choices of data are equal, we consider in this work
the problem of eddy-mean interaction in rotating stratified turbulence in the
presence of boundaries, a setting that is particularly relevant to ocean
modeling and parameterization of geostrophic mesoscale eddies. The problem
relates to the presence of rotational fluxes \cite<e.g.>{MarshallShutts81,
FoxKemper-et-al03, Maddison-et-al15}, and we provide some theoretical arguments
and evidence on why learning from the \emph{eddy force function}, which is one
method to deal with the presence of dynamically inert rotational fluxes, might
be preferable to learning from the divergence of the eddy fluxes. We will
largely leverage the experimental procedure of \citeA{BoltonZanna19}, albeit
with important differences to be detailed. While the present investigation is
largely empirical and relies on input of external knowledge that is somewhat
specific to the present problem, the present work serves to open a discussion
into data choice and/or quality, as well as probing the available information
content in data in the general case, possibly in a more systematic and objective
fashion than one performed here.

The technical problem statement relating to rotational fluxes and its plausible
impact on data quality for data-driven methods are outlined in \S\ref{sec:rot}.
In \S\ref{sec:method} we outline our experimental procedure, numerical model
used and data-driven method. \S\ref{sec:skill} summarizes the impact of data
choice on the skill of the trained models. \S\ref{sec:robust} considers the
issue of model robustness via investigating the models' skill and their
sensitivity to noise in the training data. We close in \S\ref{sec:conc} and
provide outlooks, focusing particularly on further experiments to probe the
information content of data being for use in data-driven methods of relevance to
the present eddy-mean interaction problem.

%-------------------------------------------------------------------------------

\section{Rotational fluxes and the eddy force function}\label{sec:rot}

%%%%%%%%%%%%%%%%%%%%%%%%%%%%%%%%%%%%%%%%%%%%%%

\subsection{Formulation}

For this particular work we consider turbulent motion under the influence of
strong rotation and stratification. Specifically, we consider the
Quasi-Geostrophic (QG) limit \cite<e.g.>{Vallis-GFD}, which is a widely-used and
applicable limit for oceanic mesoscale dynamics where the motion is geostrophic
at leading order. If we consider the standard Reynolds decomposition with
\begin{linenomath*}
\begin{equation}\label{eq:reynolds}
  A = \overline{A} + A', \qquad \overline{A+B} = \overline{A} + \overline{B},\quad \overline{A'} = 0, 
\end{equation}
\end{linenomath*}
where the overbar denotes a mean (with the projection operator assumed to
commute with all relevant derivatives), and a prime denotes a deviation from the
mean, the mean QG Potential Vorticity (PV) equation takes the form
\begin{linenomath*}
\begin{equation}\label{eq:qgpv}
  \frac{\partial\overline{q}}{\partial t} +
    \nabla \cdot \left( \overline{\boldsymbol{u}} \overline{q}\right) 
  = -\nabla \cdot \overline{\boldsymbol{u}' q'} + \overline{Q}.
\end{equation}
\end{linenomath*}
Here, $t$ denotes the time, $\nabla$ denotes the horizontal gradient operator,
so that the PV $q$ is defined as
\begin{linenomath*}
\begin{equation}\label{eq:pv}
  q = \nabla^2\psi + \beta y + \frac{\partial}{\partial z} \frac{f_0}{N_0^2}\frac{\partial b}{\partial z},
\end{equation}
\end{linenomath*}
where $\psi$ is the streamfunction, $f = f_0 + \beta y$ is the Coriolis
frequency (background value and leading order meridional variation), $N_0$ is
the (static) buoyancy frequency related to the imposed background
stratification, $b = f_0 \partial \psi / \partial z$ is the buoyancy,
$\boldsymbol{u} = \nabla^\perp \psi = (-\partial \psi/\partial y, \partial \psi
/ \partial x)$ is the non-divergent geostrophic velocity, and $Q$ represents all
forcing and dissipation.

An aim in studies of eddy-mean interaction is to understand the inter-dependence
of the nonlinear eddy flux terms on the right hand side of Eq.~(\ref{eq:qgpv})
and the mean state. A particular goal with eddy parameterization is to relate
the eddy flux term $\overline{\boldsymbol{u}' q'}$ with some large-scale mean
state, normally as
\begin{linenomath*}
\begin{equation}\label{eq:param_form}
  \overline{\boldsymbol{u}' q'} \sim f(\overline{q}, \ldots; \kappa, \ldots),
\end{equation}
\end{linenomath*}
where $f$ is some mapping between mean states (such as $\overline{q}$) and
associated parameters (such as $\kappa$) to the eddy fluxes. Once such a
relation exists, we take a divergence, from which we obtain the eddy forcing on
the mean. A notable example would be PV diffusion \cite<e.g.,>{Green70,
Marshall81, RhinesYoung82}, where we directly postulate for the form of $F$ as
\begin{linenomath*}
\begin{equation}\label{eq:pv_diff}
  \overline{\boldsymbol{u}' q'} = -\kappa \nabla\overline{q} \qquad \Rightarrow \qquad -\nabla\cdot\overline{\boldsymbol{u}' q'} = \nabla\cdot\left(\kappa \nabla\overline{q}\right).
\end{equation}
\end{linenomath*}
We emphasize the ordering of the operations here: we obtain a functional
relation between the mean and eddy fluxes first, then we take a divergence to
obtain the eddy forcing (cf. Fickian diffusion closures).

%%%%%%%%%%%%%%%%%%%%%%%%%%%%%%%%%%%%%%%%%%%%%%

\subsection{The issue of rotational fluxes}

The form as given in Eq.~(\ref{eq:param_form}) is suggestive that data-driven
approaches would be useful by either directly regressing/learning for the
mapping $f$, or when a mapping (cf. parameterization) such as
Eq.~(\ref{eq:pv_diff}) is given, to learn for the parameters such as $\kappa$.
However, there is a subtlety involved here, arising from the fact that it is the
divergence of the eddy fluxes that arises \cite<and is generic beyond the QG
system, where the eddy forcing arises from a divergence of the Eliassen--Palm
flux tensor, with the eddy fluxes as the tensor components, e.g.>{Young12,
MaddisonMarshall13}. A two-dimensional vector field such as
$\overline{\boldsymbol{u}' q'}$ can, via a Helmholtz decomposition, be written
as
\begin{linenomath*}
\begin{equation}\label{eq:helmholtz}
  \overline{\boldsymbol{u}' q'} = \nabla \tilde{\Psi} + \hat{\boldsymbol{e}}_z \times \nabla \tilde{\Phi} + \tilde{\boldsymbol{H}},
\end{equation}
\end{linenomath*}
where $\hat{\boldsymbol{e}}_z$ is the unit vector pointing in the vertical, and
the terms are respectively a divergent (vanishing under a curl), rotational
(vanishing under a divergence), and harmonic component (vanishing under both a
curl and divergence). Since the eddy forcing on the mean appears as a
divergence, the rotational (and harmonic) eddy fluxes are dynamically inert, and
one might expect that the presence of such dynamically inert fluxes is going to
be detrimental to the regression/learning by data-driven methods. Similar issues
arise for example arise in a diagnostic problem for the PV diffusivity $\kappa$,
where rotational fluxes are known to severely contaminate the calculation
\cite<e.g. Fig.1>{Mak-et-al16b}.

One way to get around this problem is to perform a Helmholtz decomposition as
above and only perform learning/regression/diagnoses using only the divergent
term $\nabla \tilde{\Psi}$. This approach is however complicated by the issue of
gauge freedom in the presence of boundaries \cite<e.g.,>{FoxKemper-et-al03,
Maddison-et-al15, Mak-et-al16b}. The standard Helmholtz decomposition as
commonly employed (e.g. in electromagnetism problems) is unique because we have
periodic or rapidly decaying boundary conditions. The non-uniqueness of the
Helmholtz decomposition in the presence of boundaries arises from the fact that
there is generically no inherited natural boundary condition for arbitrary
choices of vector fields (although there may be ones that are physically
relevant depending on the problem), and that the divergent term $\nabla
\tilde{\Psi}$ is unique only up to an arbitrary rotational gauge.

One possibility might be to utilize the divergence of the eddy flux directly
(e.g. $\nabla\cdot\overline{\boldsymbol{u}' q'}$). This is somewhat the approach
taken for example in the works of \citeA{BoltonZanna19} and
\citeA{ZannaBolton20}, who considers applying data-driven methods to learn about
sub-grid momentum forcing in an ocean relevant model. While they report positive
results from data-driven methods in their work, there are some points that are
worth revisiting, particularly regarding learning from the divergence of the
eddy flux. One issue is the spatial resolution of data itself: the eddy flux
data itself is already small-scale, and now we want its divergence, which is an
even finer scale quantity, so there could be sensitivity of the data to the
numerical model resolution itself. Following on from this point is the issue of
\emph{robustness}. The learning problem here is trying to find a mapping between
very small-scale data and large-scale data (e.g., divergence of eddy flux and
say some function of the streamfunction), and questions arise whether this leads
to sensitivity to the training data, or whether such a choice is unnecessarily
taxing on the machine learning algorithms. A final point is more subtle and more
speculative, to do with \emph{commutativity}, i.e. ordering of operations. Eddy
parameterizations are usually formulated as in Eq.~(\ref{eq:param_form}): we
learn a $f(\ldots) = \overline{\boldsymbol{u}' q'}$, from which we take a
divergence of the learned $f$ to get the eddy forcing. If we are learning from
$\nabla\cdot\overline{\boldsymbol{u}' q'}$, then the ordering is different,
because we are really learning for some $\nabla\cdot\overline{\boldsymbol{u}'
q'} = \hat{f}(\ldots)$, where we would hope that $\hat{f} = \nabla\cdot F$.
There is however no reason to expect such an equality in general, since the
resulting mappings $F$ or $\hat{F}$ obtained from machine learning algorithms
are nonlinear.

If we are simply interested in something that just `works', then these
aforementioned points may not actually matter. If, on the other hand, we are
interested in learning about the underlying physics via data-driven methods,
then it is not clear whether the aforementioned properties (or the lack thereof)
become fundamental limitations in the applicability of the procedure.

%%%%%%%%%%%%%%%%%%%%%%%%%%%%%%%%%%%%%%%%%%%%%%

\subsection{The eddy force function}

If we instead consider learning from data at the eddy flux level, then we
probably want to filter out the rotational component in some way, ideally in a
unique fashion. While the statement about the non-uniqueness of the Helmholtz
decomposition holds for generic tracer fluxes in the presence of boundaries, it
turns out, for the QG system and for the eddy PV flux, there is in fact a
natural boundary condition that is inherited from the no-normal flow condition
\cite{Maddison-et-al15}. The decomposition
\begin{linenomath*}
\begin{equation}\label{eq:eff}
  \overline{\boldsymbol{u}' q'} = -\nabla \Psi^q_{\rm eff} + \hat{\boldsymbol{e}}_z \times \nabla \Phi^q_{\rm eff} + \boldsymbol{H}^q,
\end{equation}
\end{linenomath*}
where $\Psi^q_{\rm eff}$ denotes the eddy force function (note the extra minus
sign on the gradient term compared to Eq.~\ref{eq:helmholtz}), and may be
obtained from solving the Poisson equation
\begin{linenomath*}
\begin{equation}\label{eq:pv_eff}
  \nabla\cdot\overline{\boldsymbol{u}' q'} = -\nabla^2 \Psi^q_{\rm eff}
\end{equation}
\end{linenomath*}
subject to homogeneous Dirichlet boundary conditions $\Psi^q_{\rm eff} = 0$.
Such an object is uniquely defined (from fixing the gauge freedom via the
naturally inherited boundary condition), and $\Psi^q_{\rm eff}$ can be proved to
be optimal in the $H^1_0$ sense, i.e. $-\nabla\Psi^q_{\rm eff}$ is a minimizer
in $L^2$, or that the dynamically active part of the eddy flux encoded by
divergent part is as `uncontaminated' as possible, at least in a simply
connected domain \cite<see Appendix A of>{Maddison-et-al15}. Furthermore, via
the linearity assumption of the eddy force function and boundary condition
inheritance \cite{Maddison-et-al15}, we can define an eddy force function for
the components that contribute towards the definition of eddy PV flux: for
example, from the definition of PV given in Eq.~(\ref{eq:pv}), we can decompose
\begin{linenomath*}
\begin{equation}\label{eq:eff_zeta}
  \overline{\boldsymbol{u}' \zeta'} = -\nabla \Psi^\zeta_{\rm eff} + \hat{\boldsymbol{e}}_z \times \nabla \Phi^\zeta_{\rm eff} + \boldsymbol{H}^\zeta,
\end{equation}
\end{linenomath*}
where $\zeta = \nabla^2\psi$ is the relative vorticity, giving rise to a
relative vorticity or momentum eddy force function $\Psi^\zeta_{\rm eff}$
\cite<related to the Reynolds stress via the Taylor identity,
e.g.>{MaddisonMarshall13}, computed via an analogous Poisson equation to
Eq.~(\ref{eq:pv_eff}) also with homogeneous Dirichlet boundary conditions, and
similarly for a buoyancy eddy force function $\Psi_{\rm eff}^b$. For
concreteness, the discussion will focus on the PV eddy force function $\Psi_{\rm
eff}^q$, but we document results from all three contributions in the later
sections.

The eddy force functions have been previously demonstrated to be a useful
quantity for diagnoses problems \cite<e.g.,>[in diagnosing eddy diffusivities
via inverse approaches]{Mak-et-al16b}, and we might expect that it would be a
useful quantity for data-driven methods applied to eddy parameterization of
rotating stratified turbulence. To compare with the discussion above, the eddy
force function is a larger-scale object, which might lead to weaker sensitivity
during the training phase compared to training on
$\nabla\cdot\overline{\boldsymbol{u}' q'}$. The gradient of the eddy force
function $-\nabla\Psi^q_{\rm eff}$ uniquely defines the dynamically relevant
eddy flux, suggesting that $-\nabla\Psi^q_{\rm eff}$ would serve as a better
choice of data compared to training on $\overline{\boldsymbol{u}' q'}$, since
the latter contains dynamically irrelevant data. Additionally, given
parameterizations are more naturally formulated as a relation between the eddy
fluxes and the mean state (cf. Eq.~\ref{eq:param_form}), $-\nabla\Psi^q_{\rm
eff}$ avoids the possible issue with commutativity mentioned above.

%-------------------------------------------------------------------------------

\section{Model details}\label{sec:method}

Taking into account the above discussion, we explore here whether the eddy force
function serves as a potentially useful object for machine learning of ocean
mesoscale turbulence. For a problem $y = f(x)$, the focus here is principally on
the skill of the models $f$, trained on various output data $y$ for the same
inputs $x$, where skill is to be measured by various mismatches between $y_{\rm
data}$ and $y_{\rm predict} = f(x_{\rm data})$. We detail here a set of
experiments to test and explore the following hypotheses:
\begin{enumerate}
  \item models trained upon the filtered eddy flux $-\nabla\Psi^q_{\rm eff}$
  would be more skillful than ones trained upon the full eddy flux
  $\overline{\boldsymbol{u}' q'}$,
  \item models trained upon the filtered eddy flux $-\nabla\Psi^q_{\rm eff}$
  would possibly be comparable in skill to ones trained upon the divergence of
  the eddy flux $\nabla\cdot\overline{\boldsymbol{u}' q'}$, but the latter
  models might be more sensitive to data quality.
\end{enumerate}

The experimental approach will largely mirror that of \citeA{BoltonZanna19}.
However, one important fundamental difference of our work is the choice of
average, which impacts the definition of eddies from Eq.~(\ref{eq:reynolds}).
Where \citeA{BoltonZanna19} take a low-pass spatial filter as the projection
operator, here we employ a time-average and has the property that $\overline{A'}
= 0$ in line with properties of a Reynolds opeartor. Our eddy forcing then is in
the more familiar form of a nonlinear eddy flux (e.g.
$\nabla\cdot\overline{\boldsymbol{u}' q'}$), rather than as a difference between
the spatially averaged quantities \cite<e.g., $\boldsymbol{S} =
\overline{\nabla\cdot(\boldsymbol{u} q)}
-\nabla\cdot(\overline{\boldsymbol{u}}\, \overline{q})$, Eq. 7
of>{BoltonZanna19}. The current definition of the eddy force function
$\Psi^q_{\rm eff}$ assumes a Reynolds average \cite{Maddison-et-al15}, and while
there are likely extensions and relaxation of assumptions possible, for
simplicity we do not pursue this avenue and utilize time-averaging.

%%%%%%%%%%%%%%

\subsection{Numerical ocean model setup}

The physical setup we consider is essentially the same three-layer QG square
double gyre configuration as \citeA{BoltonZanna19} \cite<cf.>{Berloff05a,
Karabasov-et-al09, Marshall-et-al12, Mak-et-al16b}, but solved with a
pseudo-spectral method instead of using the finite difference CABARET scheme of
\citeA{Karabasov-et-al09}. The numerical model (\verb|qgm2|) generating the data
presented in this work utilizes the parameters detailed in \citeA{Mak-et-al16b},
with the stratification parameters chosen such that the first and second Rossby
deformation radii are $32.2$ and $18.9\ \mathrm{km}$, with a horizontal grid
spacing of $\Delta x = \Delta y = 7.5\ \mathrm{km}$ (which is 512 by 512 in
horizontal grid points), a horizontal viscosity value of $\nu = 50\
\mathrm{m}^2\ \mathrm{s}^{-1}$, and a time-step of $\Delta t = 30\
\mathrm{mins}$. A wind forcing with peak wind stress of $\tau_0 = 0.8\
\mathrm{N}\ \mathrm{m}^{-2}$ is used \cite<correcting a typo in Table 1
of>{Mak-et-al16b}. The model is spun up from rest for 20,000 days, and a further
integration period of 5,000 days after this spin up is performed for computing
time-averages. 

The accumulated time-averages of the eddy fluxes are used to compute the eddy
force function $\Psi_{\rm eff}$ via solving the Poisson equation in
Eq.~(\ref{eq:pv_eff}) with homogeneous Dirichlet boundary conditions, performed
per layer. For this procedure, we leverage the FEniCS software
\cite{LoggWells10, Logg-et-al-FENICS, Alnaes-et-al14} following the previous
works of \citeA{Maddison-et-al15} and \citeA{Mak-et-al16b}, making use of the
high level abstraction, automatic code generation capabilities and the numerous
inbuilt solvers that are particularly suited to elliptic equations we have here.
The data from each grid point of the numerical model are the nodal values on a
regular structured triangular mesh, with a projection onto a piecewise linear
basis (CG1). All derivative operations are performed on the finite element mesh,
and the nodal values of the relevant fields are restructured into arrays for
feeding into the machine learning algorithms.

\begin{figure}
  \begin{center}\includegraphics[width=0.9\textwidth]{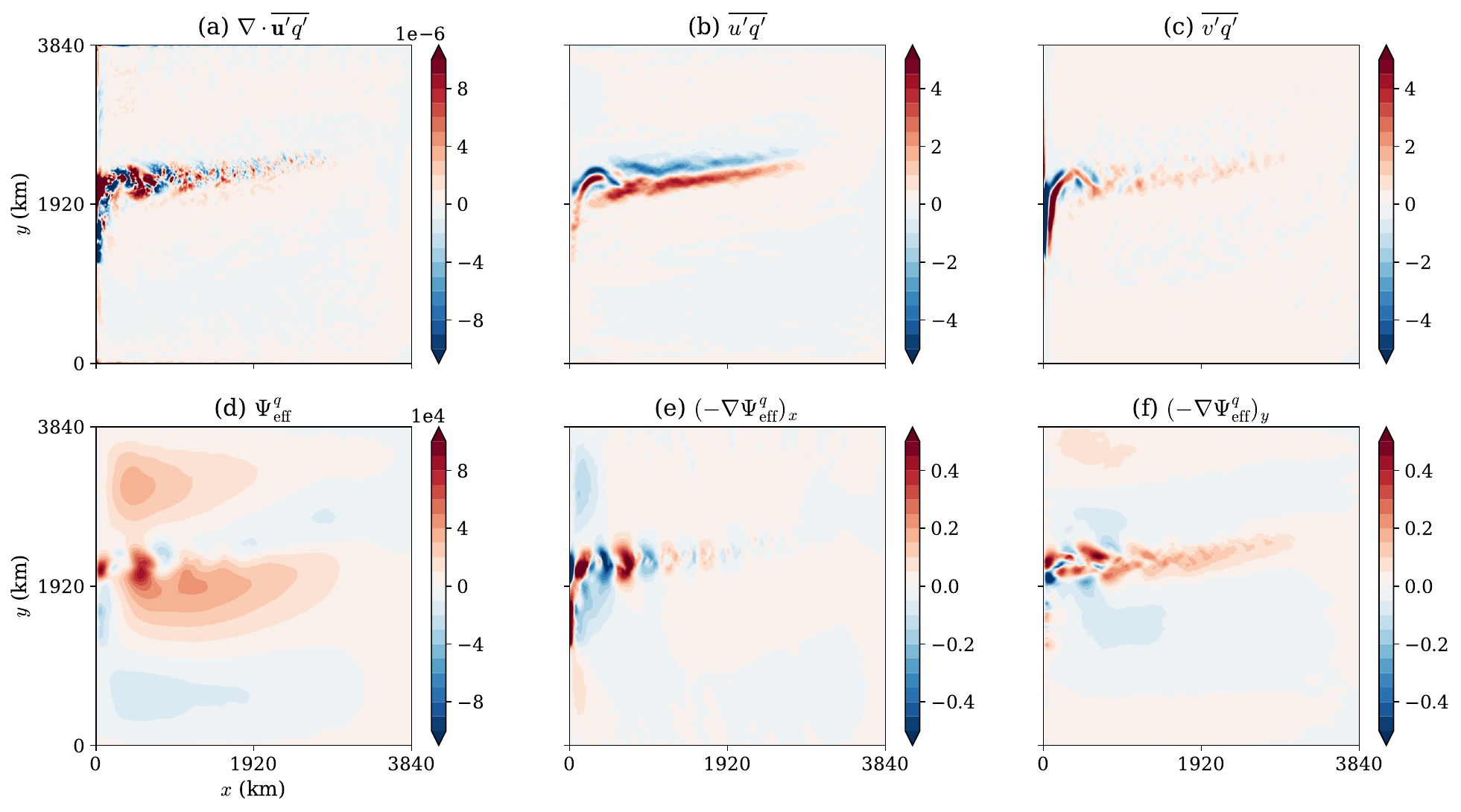}\end{center}
  \caption{($a$) The divergence of the eddy PV flux (units of
  $\mathrm{s}^{-2}$), calculated from the diagnosed time-averaged ($b$) zonal
  and ($c$) meridional component of the PV fluxes (units of $\mathrm{m}\
  \mathrm{s}^{-2}$). ($d$) The associated eddy force function $\Psi^q_{\rm eff}$
  (units of $\mathrm{m}^2\ \mathrm{s}^{-2}$) calculated from the data shown in
  panel $a$, and the ($e$) zonal and ($f$) meridional component of $-\nabla
  \Psi^q_{\rm eff}$, the associated eddy PV fluxes with the dynamically inert
  rotational component removed (units of $\mathrm{m}\ \mathrm{s}^{-2}$). Note
  the different choices of colorbar limits between the data range in panels
  ($b,c$) and ($e,f$).}
  \label{fig:f1}
\end{figure}

Fig.~\ref{fig:f1} shows some sample output data in the surface layer. The two
horizontal components of the time-averaged eddy PV fluxes in panels ($b,c$) are
the datasets returned by the numerical model, which is sampled onto a finite
element mesh as a vector object. The resulting object's divergence can then be
computed, and the result is given in panel ($a$). As expected, the divergence of
the eddy PV flux has more smaller-scale fluctuations and is less smooth than the
eddy PV fluxes. Solving the relevant Poisson equation in FEniCS, the PV eddy
force function $\Psi_{\rm eff}^q$ is shown in panel ($d$). From
\citeA{Maddison-et-al15}, the gradient of the eddy force function
$\nabla\Psi^q_{\rm eff}$ has a physical interpretation when considered together
with the time-mean streamfunction $\overline{\psi}$ \cite<not shown, but
see>{Maddison-et-al15}, interpreted as whether eddies are accelerating the
mean-flow (if $\nabla\Psi^q_{\rm eff} \cdot \nabla \overline{\psi} > 0$,
interpreted as an input of energy into the mean by eddies) or decelerating the
mean flow (if $\nabla\Psi^q_{\rm eff} \cdot \nabla \overline{\psi} < 0$,
interpreted as an extraction of energy from the mean by eddies). Here, the eddy
force function can be shown to correspond to the regimes where the eddies are
slowing down the mean-flow via baroclinic instability when the Western Boundary
Current first separates (the first positive-negative pattern emanating from the
western boundary, which is anti-correlated with $\nabla\overline{\psi}$), while
the next dipole pattern (the first negative-positive patterns, which is
correlated with $\nabla\overline{\psi}$) is an eddy forcing of the mean-flow
corresponding to an eddy driven regime \cite<cf.>{WatermanJayne11,
WatermanHoskins13}.

From this $\Psi_{\rm eff}^q$, the horizontal components of the gradient leads to
the eddy PV fluxes with the rotational component removed, and are shown in
panels ($e,f$). While not obvious at first sight, the divergence of the full
eddy PV flux (panels $b,c$) and the divergence of the filtered eddy PV flux
(panels $e,f$) are both equal to $\nabla\cdot\overline{\boldsymbol{u}'q'}$
(panel $a$) up to numerical solver errors (here at least four orders of
magnitude smaller than the data). In this instance, note also that the filtered
eddy flux has qualitatively different spatial patterns to the full eddy flux,
and that the filtered eddy flux is around an order of magnitude smaller than the
full eddy fluxes. The behavior is consistent with observations that the
rotational eddy fluxes can be large \cite<e.g.>{Griesel-et-al09}, and suggests
we probably do want to filter the dynamically inert component out should we
utilize eddy flux data to learn about geostrophic turbulence.

%%%%%%%%%%%%%%

\subsection{Model training procedure}

Following \citeA{BoltonZanna19} we employ Convolutional Neural Networks
\cite<CNNs; e.g., \S9,>{Goodfellow-et-al-16} to map between the specified inputs
and targets. In line with the intended investigation, the choice of parameters
for training the CNNs are kept fixed and chosen as in \citeA{BoltonZanna19}, and
the main quantity we vary is the choice of output data. The mappings that are
returned as a CNN are denoted:
\begin{itemize}
  \item $f_{\rm div}^q(\ldots)$, with output data as the divergence of the eddy PV flux
  $\nabla\cdot\overline{\boldsymbol{u}'q'}$,
  \item $f_{\rm full}^q(\ldots)$, with output target data as the full eddy PV flux
  $\overline{\boldsymbol{u}'q'}$,
  \item $f_{\rm eff}^q(\ldots)$, with output data as the dynamically active eddy
  PV flux as defined through a gradient of the PV eddy force function (cf.
  Eq.~\ref{eq:pv_eff}) $-\nabla\Psi^q_{\rm eff}$.
\end{itemize}
Note that $f_{\rm div}^q(\ldots)$ predicts a scalar field, while the $f_{\rm
full/eff}^q(\ldots)$ returns a vector field. A possible choice could be to train
a model from the eddy force function, and from the trained model's predicted
eddy force function compute its Laplacian to obtain the divergence of the eddy
flux. As mentioned above, this is an extremely difficult test for model skill
since gradient operations amplify mismatches, and we comment on related results
and observations are in the conclusions section.

\begin{figure}
  \begin{center}\includegraphics[width=0.9\textwidth]{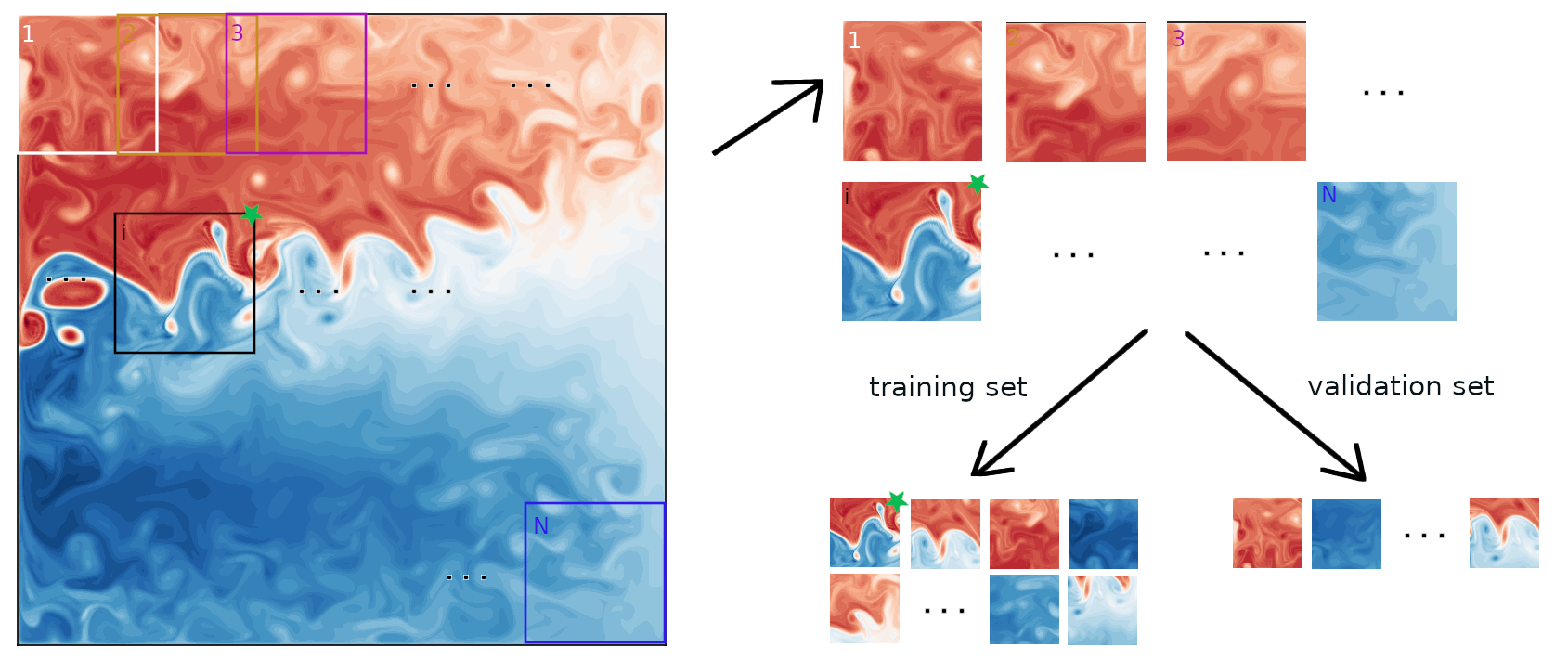}\end{center}
  \caption{Model training strategy demonstrated here with a snapshot of the
  instantaneous PV from model output. The domain is partitioned into small
  square regions of size 40 by 40 pixels, overlapping in the $x$ and $y$
  direction by 6 pixels, resulting in 6400 entries of input and output data.
  Each pair of input and output data is assigned with equal probability to be in
  the training set and validation set at the 80:20 ratio, from which a trained
  model results. An ensemble of models with 20 members is created, and are
  tested according to the procedure detailed in text.}
  \label{fig:f2}
\end{figure}

To train up these mappings in the present time-averaged case, we follow the
schematic given in Fig.~\ref{fig:f2}, partially inspired by the approach of
\citeA{BoltonZanna19}. The model domain is partitioned into small overlapping
boxes. The input and output data associated with each of these boxes are paired
up, and the pairs are each assigned a number and randomly shuffled (i.e.
sampling from a uniform probability distribution function) depending on a choice
of a random seed, and subsequently assigned to the training set (for training up
the model) and validation (for tuning the hyperparameters in order to minimize a
specified loss function) set with a 80:20 ratio. A model is trained up, and the
skill of the model is its ability to be able to predict the global field. In the
512 by 512 pixel domain, we take the small boxes to be 40 by 40 pixels, with a
stride of six, resulting in a collection of $80^2 = 6400$ images of the domain.
For statistical significance, an ensemble of 20 such models were trained up,
each ensemble member only differing in the choice of the random seed, and the
same sets of random seeds are used for the ensembles to be compared against. The
CNNs are built using the PyTorch platform \cite{Paszke-et-al19}, where the CNN
architecture consists of three hidden convolutional layers with square kernels
(of size 8, 4 and 4 respectively), with a two-dimension max pooling layer with
square kernel of size 2, and a fully-connected linear activation layer as the
output. The CNNs are trained with a batch size of 64, using the Adam optimizer
\cite{KingmaBa14} with a mean squared error loss function. An early stopping
criterion is used to monitor the loss function during the training to avoid
over-fitting; for simplification, we use a constant learning rate of $10^{-4}$
during training.

%-------------------------------------------------------------------------------

\section{Model skill}\label{sec:skill}

We first evaluate the predictive skill of the various models to the choice of
target data. The skill of the models are judged by its ability to reduce
mismatches of the divergence of the eddy PV flux, via repeated predictions of
smaller patches (here taken with a stride of 2 pixels), with averages taken as
necessary. Note that while $f_{\rm div}^q(\ldots)$ already predicts the
divergence of the eddy PV flux, we will take a divergence of the outcome of
$f_{\rm full/eff}^q(\ldots)$ to give the predicted divergence of the eddy PV
flux. The normalized mismatch between data and prediction will be judged as
\begin{linenomath*}
\begin{equation}\label{eq:l2_norm}
  \epsilon^q_{L^2}(F^q_{(\cdot)}) = \frac{\|\nabla\cdot\overline{\boldsymbol{u}'q'} - F^q_{(\cdot)}(\ldots)\|^2_{L^2}}{\|\nabla\cdot\overline{\boldsymbol{u}'q'}\|^2_{L^2}},
\end{equation}
\end{linenomath*}
where $F_{(\cdot)}^q$ denotes the divergence of the eddy PV flux from the models
$f^q_{(\cdot)}(\ldots)$, and the $L^2$ norm is defined as
\begin{linenomath*}
\begin{equation}\label{eq:l2}
  \|g\|^2_{L^2} = \int_A g^2\; \mathrm{d}A
\end{equation}
\end{linenomath*}
for some scalar field $g$. Each ensemble member will make a set of predictions
with an associated mismatch, and the associated averages and standard deviations
computed to gauge model skill.

We note that the test for skill chosen here is inherently harder and biased
\emph{against} the models trained on the eddy PV fluxes (filtered or otherwise),
since an extra divergence operation is required in computing the mismatches. The
above choice to compare the divergence of the eddy PV flux was taken noting that
we want a quantity that is comparable across the three sets of models, and there
is a theoretical issue in comparing quantities at the eddy PV flux level (since
that requires integrating the prediction of $F^q_{\rm div}(\ldots)$, which is
then subject to a choice of boundary condition). One could argue whether it is
the $L^2$ mismatches we are ultimately interested in, since we may for example
be interested in the patterns of the forcing, rather than the exact locations of
the forcing. As a compromise, we consider the Sobolev semi-norms
\cite<e.g.>{Thiffeault12} given by
\begin{linenomath*}
\begin{equation}\label{eq:sobolev}
  \|g\|^2_{\dot{H}^{p}} = \int_A |(-\nabla^2)^{p} g|^2\; \mathrm{d}A = \sum_{k^2 + l^2\neq 0} (k^2 + l^2)^{p}|\hat{g}_{k,l}|^2,
\end{equation}
\end{linenomath*}
where $\hat{g}_{k,l}$ are the Fourier coefficients of $g$, $(k,l)$ are the
respective wavenumbers, and the link between integral and sum follows from
Parseval's theorem (e.g. if $p=0$ then it is the $L^2$ norm above when the
$k=l=0$ mode is included). Sobolev semi-norms with negative $p$ will
weigh the lower wavenumbers (i.e. the larger-scale patterns) more, and in this
instance a lower normalized mismatch
\begin{linenomath*}
\begin{equation}\label{eq:hp_norm}
  \epsilon^q_{\dot{H}^{p}}(F^q_{(\cdot)}) = \frac{\|\nabla\cdot\overline{\boldsymbol{u}'q'} - F^q_{(\cdot)}(\ldots)\|^2_{\dot{H}^{p}}}{\|\nabla\cdot\overline{\boldsymbol{u}'q'}\|^2_{\dot{H}^{p}}}
\end{equation}
\end{linenomath*}
indicates that the mismatches at the large-scales are smaller. Since we are
dealing with finite approximations so that $k^2 + l^2 < \infty$, we can perform
the computation, although the formal definition for the $\dot{H}^{p}$ semi-norms
is generally for fields with zero mean and on a periodic domain and such that the
infinite sum converges. For the work here we will focus on the case of $p=-1/2$,
sometimes referred to as the mix-norm \cite<e.g.>{Thiffeault12}; conclusions
below are qualitatiely the same if $p=-1$ or $p=-2$ were chosen (not shown).

%%%%%%%%%%%%%%

\subsection{Models trained on eddy PV fluxes}\label{sec:pv}

\begin{figure}
  \begin{center}\includegraphics[width=0.9\textwidth]{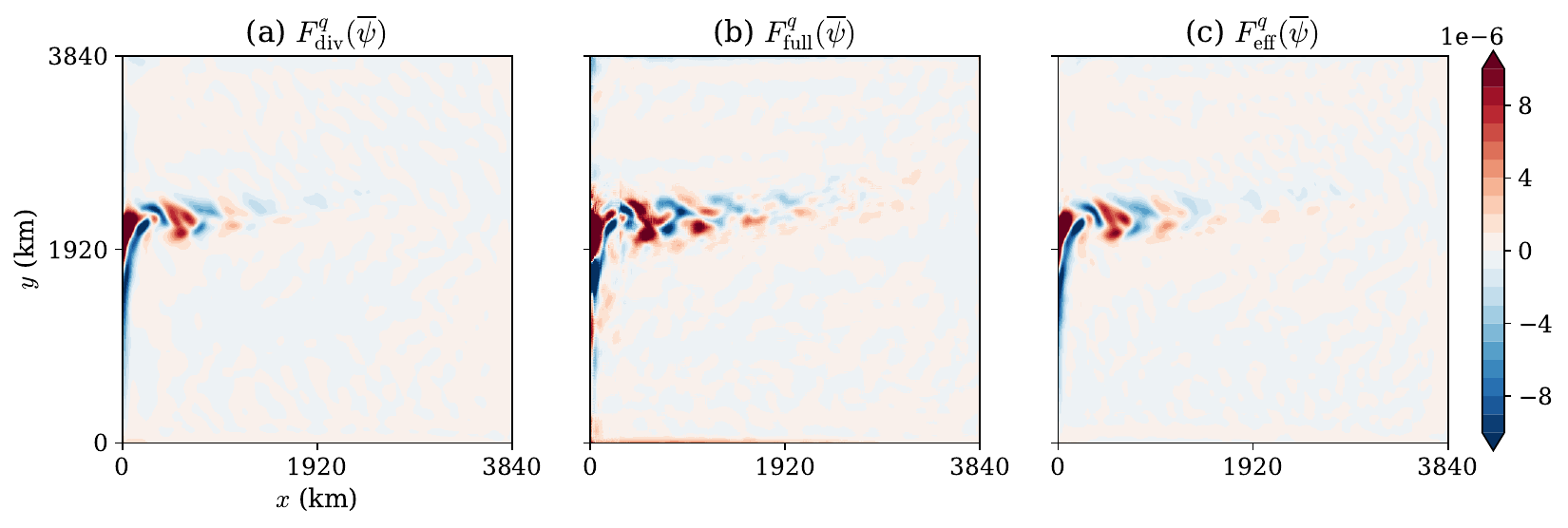}\end{center}
  \caption{Prediction of the divergence of eddy PV flux (units of $\mathrm{m}^2\
  \mathrm{s}^{-2}$) from one of the ensemble member of models. ($a$) $F^q_{\rm
  div}(\overline{\psi})$, ($b$) $F^q_{\rm full}(\overline{\psi})$, ($c$)
  $F^q_{\rm eff}(\overline{\psi})$. The target reference data shown in
  Fig.~\ref{fig:f1}$a$.}
  \label{fig:f3}
\end{figure}

We first focus on models trained up on the data based on the eddy PV flux
$\overline{\boldsymbol{u}'q'}$ with the time-mean streamfunction
$\overline{\psi}$ as the input. Fig.~\ref{fig:f3} shows the predicted divergence
of the eddy PV flux $F_{\rm div/full/eff}^q(\overline{\psi})$ as an output from
one of the model ensemble members. Compared to the target given in
Fig.~\ref{fig:f1}($a$), the predictions are more smooth with fewer small-scale
features, arising from a combination of the fact that CNNs were used, and that
our prediction step leads to some averaging of the overlaping regions. Visually,
the predictions $F_{\rm div}^q(\overline{\psi})$ and $F_{\rm
eff}^q(\overline{\psi})$ are almost indistinguishable (the latter having a
slightly stronger signal downstream of the Western Boundary Current). On the
other hand, the prediction $F_{\rm full}^q(\overline{\psi})$ shows more
fluctuation features than the other two cases. The larger amount of small-scale
features in $F_{\rm full}^q(\overline{\psi})$ likely arises because the model is
predicting the eddy PV flux first, before taking a numerical divergence of the
data, so any small fluctuations that arise from the prediction is amplified by
the divergence operation. In that regard, the fact that the prediction $F_{\rm
eff}^q(\overline{\psi})$ is so similar to $F_{\rm div}^q(\overline{\psi})$ is
rather remarkable.

\begin{figure}
  \begin{center}\includegraphics[width=\textwidth]{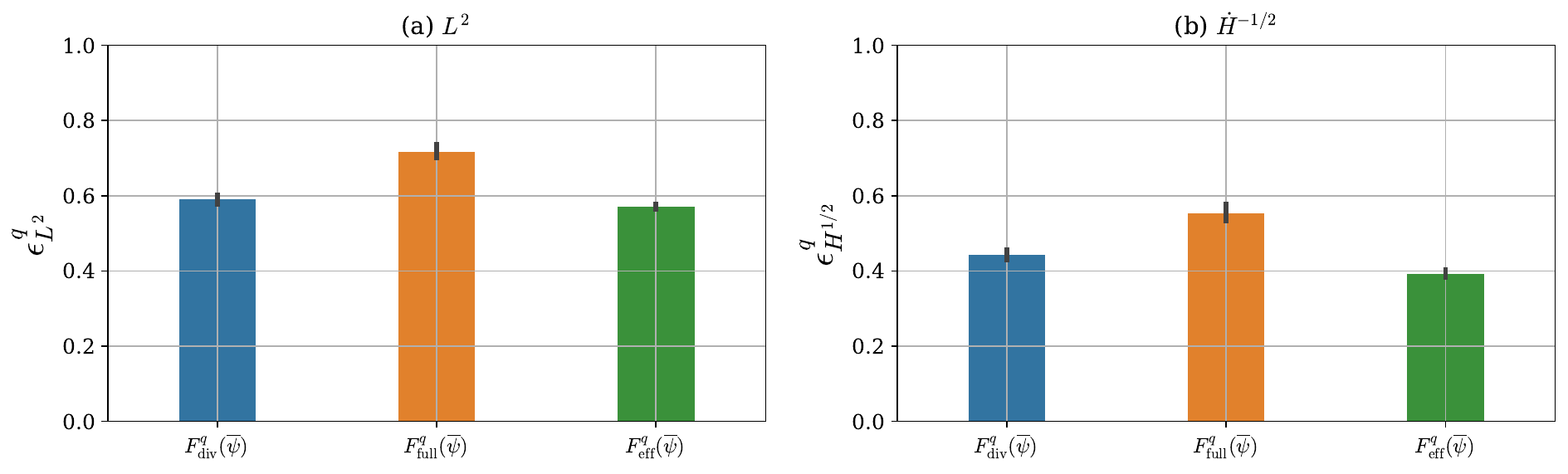}\end{center}
  \caption{Ensemble average and quartiles of the mismatch as measured by the
  normalized ($a$) $L^2$ norm ($b$) $\dot{H}^{-1/2}$ semi-norm, given by
  Eq.~(\ref{eq:l2_norm}) and (\ref{eq:hp_norm}) respectively, for the models
  predicting the divergence of the eddy PV flux (Fig.~\ref{fig:f1}$a$). Blue
  denotes models trained on the divergence of the eddy fluxes, orange denotes
  models trained on the full eddy fluxes, and green denotes models trained on
  the filtered eddy fluxes.}
  \label{fig:f4}
\end{figure}

Fig.~\ref{fig:f4} shows the more quantitative measure of computing various
mismatches in the $L^2$ norm and the $\dot{H}^{-1/2}$ semi-norm given in
Eq.~(\ref{eq:l2}) and (\ref{eq:sobolev}) respectively. The results show that the
models trained upon the filtered eddy PV flux $-\nabla\Psi_{\rm eff}^q$
outperforms the models trained upon the full eddy PV flux
$\overline{\boldsymbol{u}'q'}$, and have a comparable or even better performance
compared to the models trained up on the divergence of the eddy PV flux
$\nabla\cdot\overline{\boldsymbol{u}'q'}$. The differences in skill are visually
obvious between the models trained on the full eddy flux
$\overline{\boldsymbol{u}'q'}$ and the filtered eddy flux $-\nabla\Psi_{\rm
eff}^q$. The difference between the models trained from the filtered eddy flux
$-\nabla\Psi_{\rm eff}^q$ and the divergence of the eddy flux
$\nabla\cdot\overline{\boldsymbol{u}'q'}$, while notable in the $\dot{H}^{-1/2}$
measure, is too close to call in the $L^2$ measure (e.g. we do not have $p<0.05$
using the Student's $t$-test \cite{Student08} under the null hypothesis that the
means of $F_{\rm div}^q(\overline{\psi})$ and $F_{\rm eff}^q(\overline{\psi})$
are the same).

The results here lend support to our expectation that the presence of rotational
fluxes contaminate and degrade the accuracy of a trained up model, and that the
eddy force function provides an viable alternative for use in machine learning
approaches that addresses the problem of dynamically inert rotational fluxes,
leading to at least comparable performance from a skill point of view (and some
evidence to suggest it might be better, although that is dependent on the choice
of metric). The observation that $F_{\rm eff}^q(\overline{\psi})$ is comparable
to $F_{\rm div}^q(\overline{\psi})$ is all the more remarkable when we note that
tests based on the models' ability in reproducing the divergence of the eddy
flux is intrinsically harder and biased against models trained on
$-\nabla\Psi_{\rm eff}^q$, since an additional divergence operation that is
expected to amplify errors is required to produce $F_{\rm
eff}^q(\overline{\psi})$.

%%%%%%%%%%%%%%

\subsection{Other choice of eddy fluxes and inputs}

\begin{figure}
  \begin{center}\includegraphics[width=0.9\textwidth]{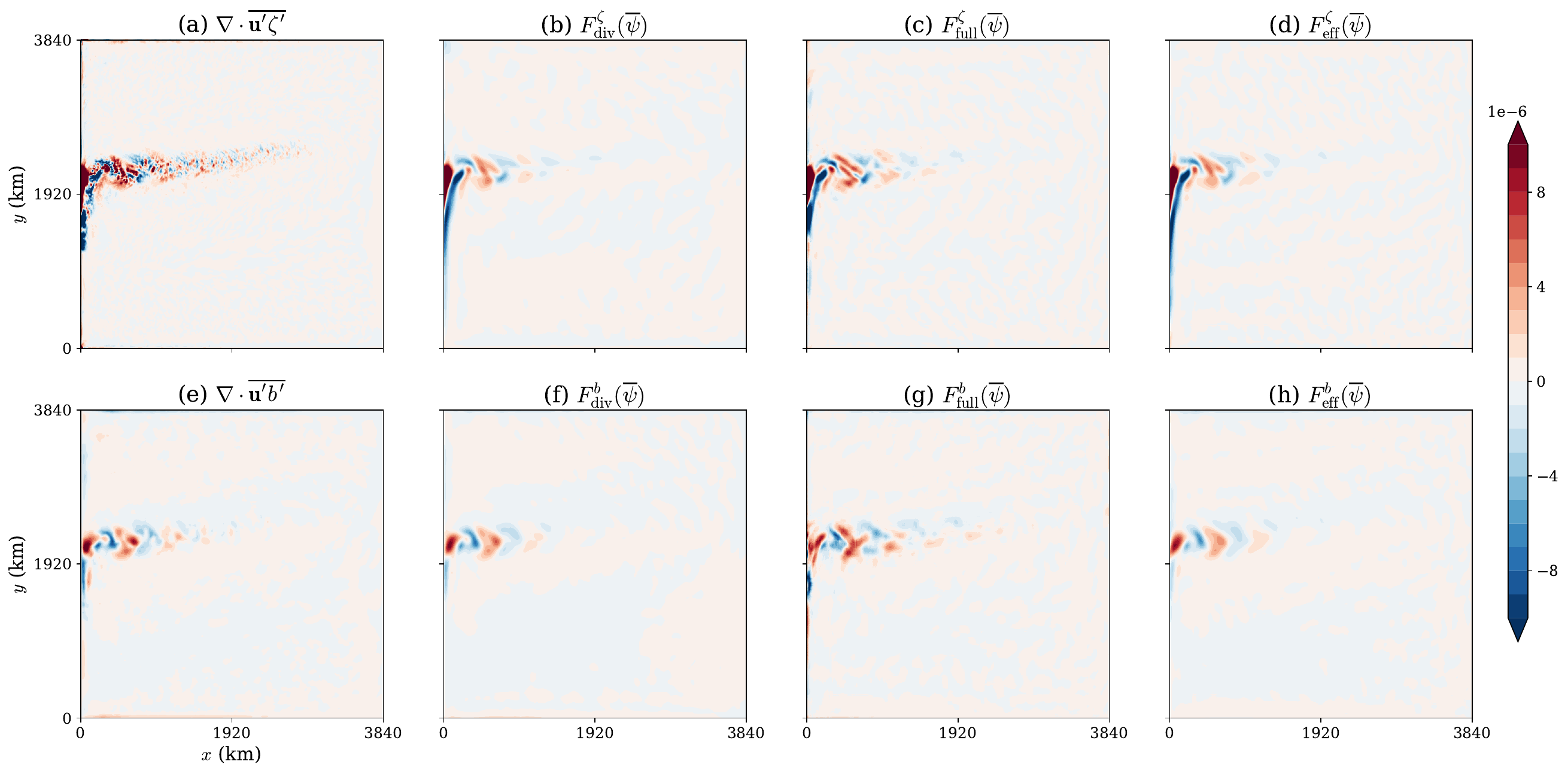}\end{center}
  \caption{Target data and predictions associated with (top row) eddy relative
  vorticity flux (related to the Reynolds stress, units of $\mathrm{m}\
  \mathrm{s}^{-2}$) and (bottom row) eddy buoyancy flux (related to the form
  stress, units also of $\mathrm{m}\ \mathrm{s}^{-2}$ taking into account of the
  extra factors). Showing ($a,e$) the divergence of the time-averaged eddy
  relative vorticity and buoyancy flux, and a sample ($b,f$) $F^{\zeta/b}_{\rm
  div}(\overline{\psi})$, ($c,g$) $F^{\zeta/b}_{\rm full}(\overline{\psi})$,
  ($d,h$) $F^{\zeta/b}_{\rm eff}(\overline{\psi})$ from one of the ensemble
  members.}
  \label{fig:f5}
\end{figure}

By the linearity assumption in deriving the eddy force function and the
definition of PV, analogous eddy force functions for momentum and buoyancy may
be defined by a similar decomposition but using the eddy relative vorticity flux
$\overline{\boldsymbol{u}'\zeta'}$ (related to the Reynolds stress via the
Taylor identity) and $\overline{\boldsymbol{u}'b'}$ (related to the form
stress). Following the notation outline above, Fig.~\ref{fig:f5} show the target
data $\nabla\cdot\overline{\boldsymbol{u}'\zeta'}$ and
$\nabla\cdot\overline{\boldsymbol{u}'b'}$, and the analogous predictions of the
divergence of the fluxes denoted by $F_{\rm
div/full/eff}^{\zeta/b}(\overline{\psi})$.

For the models trained on the data relating to the eddy PV flux shown in
Fig.~\ref{fig:f1}, the predictions are more smooth than the diagnosed target
data, and is particularly noticeable for prediction of the divergence of the
eddy relative vorticity flux in Fig.~\ref{fig:f5}($b,c,d$). For the eddy
buoyancy case, the diagnosed target data is already relatively smooth. We note
that, visually, $F^b_{\rm full}(\overline{\psi})$ in Fig.~\ref{fig:f5}($g$)
seems to be possess extra features particularly in the downstream region, while
$F^b_{\rm eff}(\overline{\psi})$ and $F^b_{\rm div}(\overline{\psi})$ in
Fig.~\ref{fig:f5}($f,h$) seems to be capturing the patterns in the target data
well, with some visual hints that the prediction from $F^b_{\rm
div}(\overline{\psi})$ has slightly sharper features.

\begin{figure}
  \begin{center}\includegraphics[width=\textwidth]{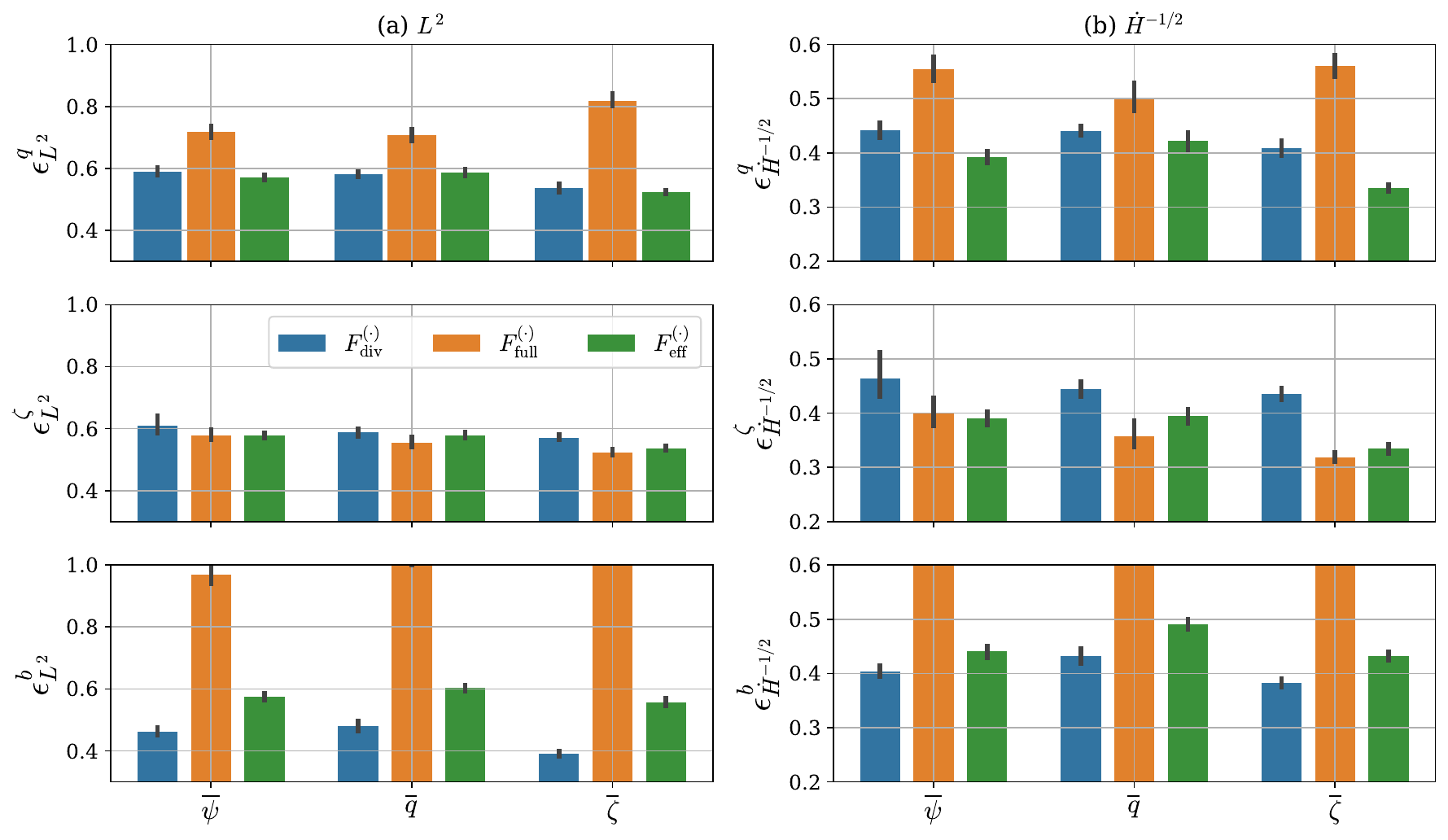}\end{center}
  \caption{Ensemble average and quartiles of the normalized ($a$) $L^2$ norm
  ($b$) $\dot{H}^{-1/2}$ semi-norm, given by Eq.~(\ref{eq:l2_norm}) and
  (\ref{eq:hp_norm}) respectively, for the models predicting the divergence of
  the eddy (rows) PV flux, relative vorticity (cf. momentum) flux, and buoyancy
  flux, over various choices of inputs. Blue denotes models trained on the
  divergence of the eddy fluxes, orange denotes models trained on the full eddy
  fluxes, and green denotes models trained on the filtered eddy fluxes. Top row
  is identical to Fig.~\ref{fig:f4}. The mismatches in $F^b_{\rm
  full}(\overline{q})$ and $F^b_{\rm full}(\overline{\zeta})$ are out of range
  of panel, with values around 1.0 and 1.5 respectively.}
  \label{fig:f6}
\end{figure}

For a more quantitative measure, we show in Fig.~\ref{fig:f6} the $L^2$ and
$\dot{H}^{-1/2}$ mismatches in $F^{q/\zeta/b}_{\rm div / full /
eff}(\overline{\psi} / \overline{q} / \overline{\zeta})$, totaling the $3^3 =
27$ possible combinations. The conclusions over all these possible choices are
largely what was drawn from before but with minor differences. The models
trained up on the filtered eddy fluxes outperform those trained upon the full
eddy fluxes (except for the case of eddy relative vorticity fluxes), and are
comparable or better than models trained on the divergence of the flux (except
in the case of the eddy buoyancy fluxes).

Noting that eddy PV fluxes have contributions from the eddy buoyancy as well as
eddy relative vorticity fluxes, it is curious that while models trained on the
filtered eddy fluxes compared with models trained on the divergence of the flux
appear to perform worse for the eddy buoyancy flux (bottom row of
Fig.~\ref{fig:f6}), but has reasonable performance in the eddy relative
vorticity flux case (middle row of Fig.~\ref{fig:f6}) such that, together, the
resulting skill in the eddy PV flux (top row of Fig.~\ref{fig:f6}) still remains
comparable (and possibly slightly better in the $\dot{H}^{-1/2}$ semi-norm,
indicating better matching in terms of large-scale patterns). One possible
explanation for the degradation in performance for eddy buoyancy fluxes is that
$\nabla\cdot\overline{\boldsymbol{u}'b'}$ is already relatively smooth and
larger-scale (Fig.~\ref{fig:f5}$e$), which might be favorable for direct use as
training data. On the other hand, the eddy relative vorticity fluxes are
inherently smaller-scale (Fig.~\ref{fig:f5}$a$), and the presence of small-scale
fluctuation might be unfavorable for direct use as training data, but does not
affect models trained on the filtered fluxes as such since the training data is
by definition more smooth. The performance of models based on the full eddy
relative vorticity fluxes is somewhat surprising, but may be to do with the
smaller component of the rotational fluxes: examining the decomposition into
divergent and rotational parts via the eddy force function (cf.
Fig.~\ref{fig:f1}$b,c,e,f$, not shown) it is found that the divergent component
is smaller by about a factor of 2 in the eddy relative vorticity flux, but a
factor of 10 in the eddy buoyancy and PV flux. The results seem to suggest that
the main benefits of filtering dynamically inert rotational fluxes would be in
the eddy buoyancy and PV.

\begin{figure}
  \begin{center}\includegraphics[width=0.9\textwidth]{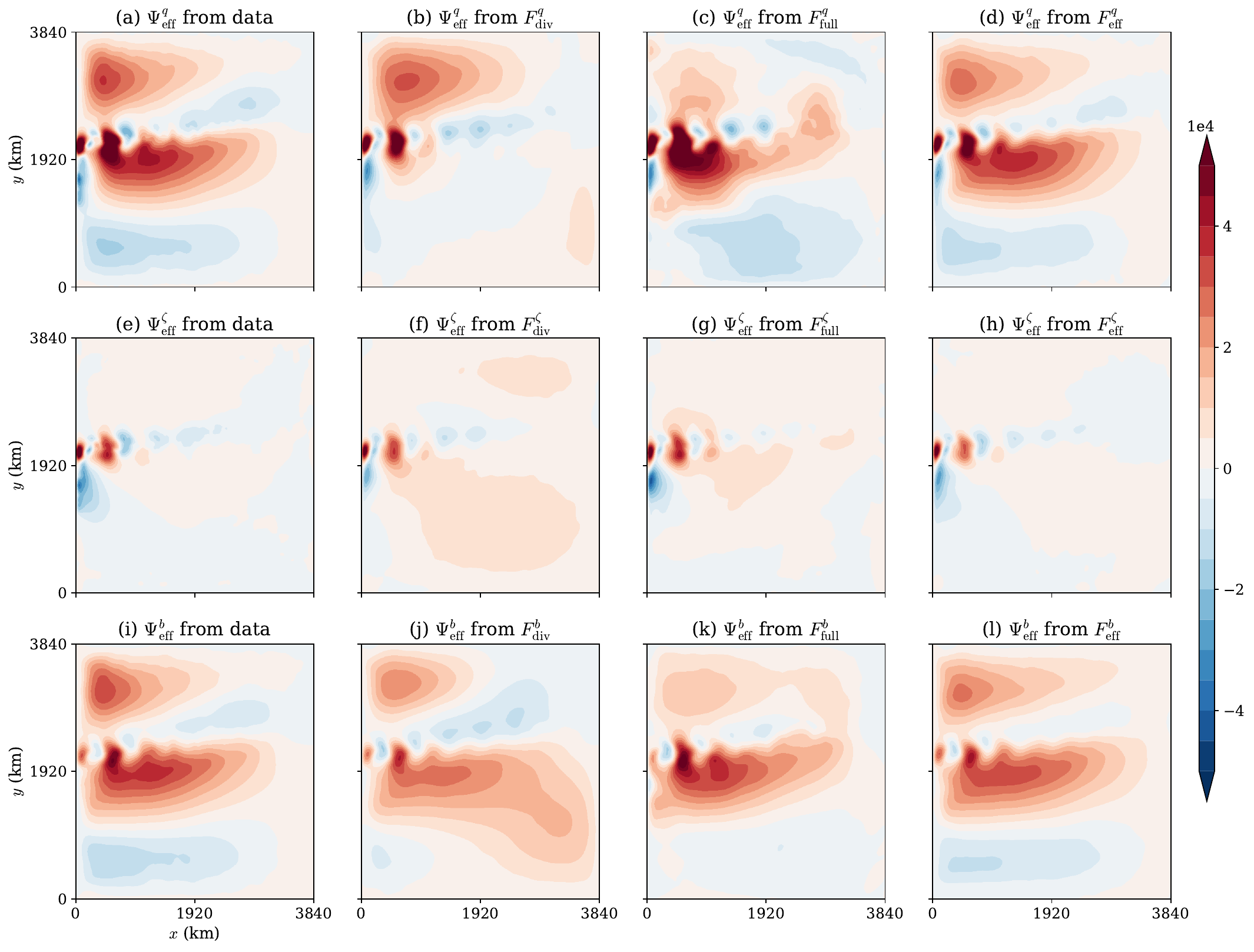}\end{center}
  \caption{($a,e,i$) Target eddy force functions $\Psi_{\rm eff}^{q/\zeta/b}$,
  and eddy force functions associated with prediction from ($b,f,j$) divergence
  of the eddy PV, relative vorticity and buoyancy fluxes, ($c,g,k$) full eddy
  PV, relative vorticity and buoyancy fluxes, and ($d,h,i$) filtered eddy PV,
  relative vorticity and buoyancy fluxes, from one of the ensemble members. All
  data shown here are in units of $\mathrm{m}^2\ \mathrm{s}^{-2}$.}
  \label{fig:f7}
\end{figure}

For completeness, we show in Fig.~\ref{fig:f7} the analogous eddy force
functions associated with the predictions from the trained models from one of
the ensemble members (although observations detailed here are robust upon
examining the outputs from other members); note the appropriate mismatches would
be closely related to the $\dot{H}^{-2}$ semi-norm as defined in
Eq.~(\ref{eq:sobolev}), but with a difference in the boundary conditions. The
predictions from models trained on the filtered eddy fluxes (panels $d,h,i$)
have patterns that are largely aligned with the diagnosed eddy force functions
from the data (panels $a,e,i$) up to minor discrepancies (e.g. downstream
patterns in panel $d$ compared to panel $a$, and panel $l$ compared to panel
$i$). The predictions from models trained on the full eddy fluxes (panels
$c,g,k$) show similar patterns although with somewhat more mismatches,
particularly in the PV and buoyancy eddy force functions. By contrast, the
predictions from the divergence of the eddy fluxes (panels $b,f,j$) show
large-scale disagreements in all three variables, the mismatches being visually
the gravest in the PV and buoyancy variables. Given that the eddy force function
encodes the dynamically active eddy fluxes, and has an interpretation that
$\nabla \Psi_{\rm eff} \cdot \nabla \overline{\psi}$ encodes the sign of energy
exchange between the mean and eddy component \cite{Maddison-et-al15}, the
finding here suggests the predictions from models trained on the divergence of
the eddy fluxes are very likely representing erroneous energy transfers,
particularly for processes associated with eddy buoyancy fluxes.

%-------------------------------------------------------------------------------

\section{Model robustness}\label{sec:robust}

The above observations of model skill and its sensitivity to small-scale
fluctuations brings into question the issue of robustness particularly for the
models trained on the divergence of the eddy fluxes. To explore the sensitivity
of skill to noise in the data, we consider a set of experiments where we add
noise $\eta(x,y)$ to the data at the \emph{training} stage, and judge the
models' performance on its ability in predicting the target data without noise.
To make sure we are comparing models in a consistent manner, we add an
appropriately scaled Gaussian distributed noise $\eta(x,y)$ to the eddy fluxes
($\overline{\boldsymbol{u}'q'}, \overline{\boldsymbol{u}'\zeta'},
\overline{\boldsymbol{u}'b'}$), from which we compute the divergence of the eddy
flux as well as the eddy force function from the noisy data, and train up the
models using the procedure outlined above. In that sense the whole set of models
are exposed to the \emph{same} choice of noise, since 1 unit of noise at the
divergence level is not necessarily the same as 1 unit of noise at the
streamfunction level. The noise level here is measured in units of the standard
deviation of the eddy flux data. The hypothesis is that the models trained on
the filtered eddy fluxes are more robust than those trained on the divergence of
the eddy fluxes, and able to maintain model skill with increased levels of
noise.

A note to make here is that the stochastic noise $\eta(x,y)$ in this regard is
formally non-differentiable in space, so that the divergence operation on it is
not well-defined. In terms of numerical implementation, however, the random
numbers sampled from the appropriately scaled Gaussian distribution are the
nodal values of the finite element mesh used in FEniCS, and there is a
projection onto a linear basis, so that a derivative operation on the projected
$\eta(x,y)$ is allowed within FEniCS, though the operation may be numerically
sensitive. An approach we considered is filtering the noise field. We consider
solving for some $\tilde{\eta}(x,y)$ satisfying
\begin{linenomath*}
\begin{equation}\label{eq:helmholtzM}
  (1 - L^2\nabla^2)^2\tilde{\eta} = \eta
\end{equation}
\end{linenomath*}
with no-flux boundary conditions, and it is the resulting $\tilde{\eta}(x,y)$
that is added to the training data. The resulting $\tilde{\eta}$ is by
construction differentiable at least once so that a divergence is well-defined.
For the operator $(1 - L^2\nabla^2)^2$, the associated Green's function has a
characteristic length-scale $L$ that can be interpreted as a filtering
length-scale where the radial spectral power density decreases significantly
after $L$ \cite<closely related to the Mat\'ern auto-covariance,
e.g.>{Whittle63, Lindgren-et-al11}. Note that `noise level' here refers to the
magnitude of $\eta(x,y)$, and that $\max|\tilde{\eta}(x,y)| < \max|\eta(x,y)|$
by construction.

\begin{figure}
  \begin{center}\includegraphics[width=\textwidth]{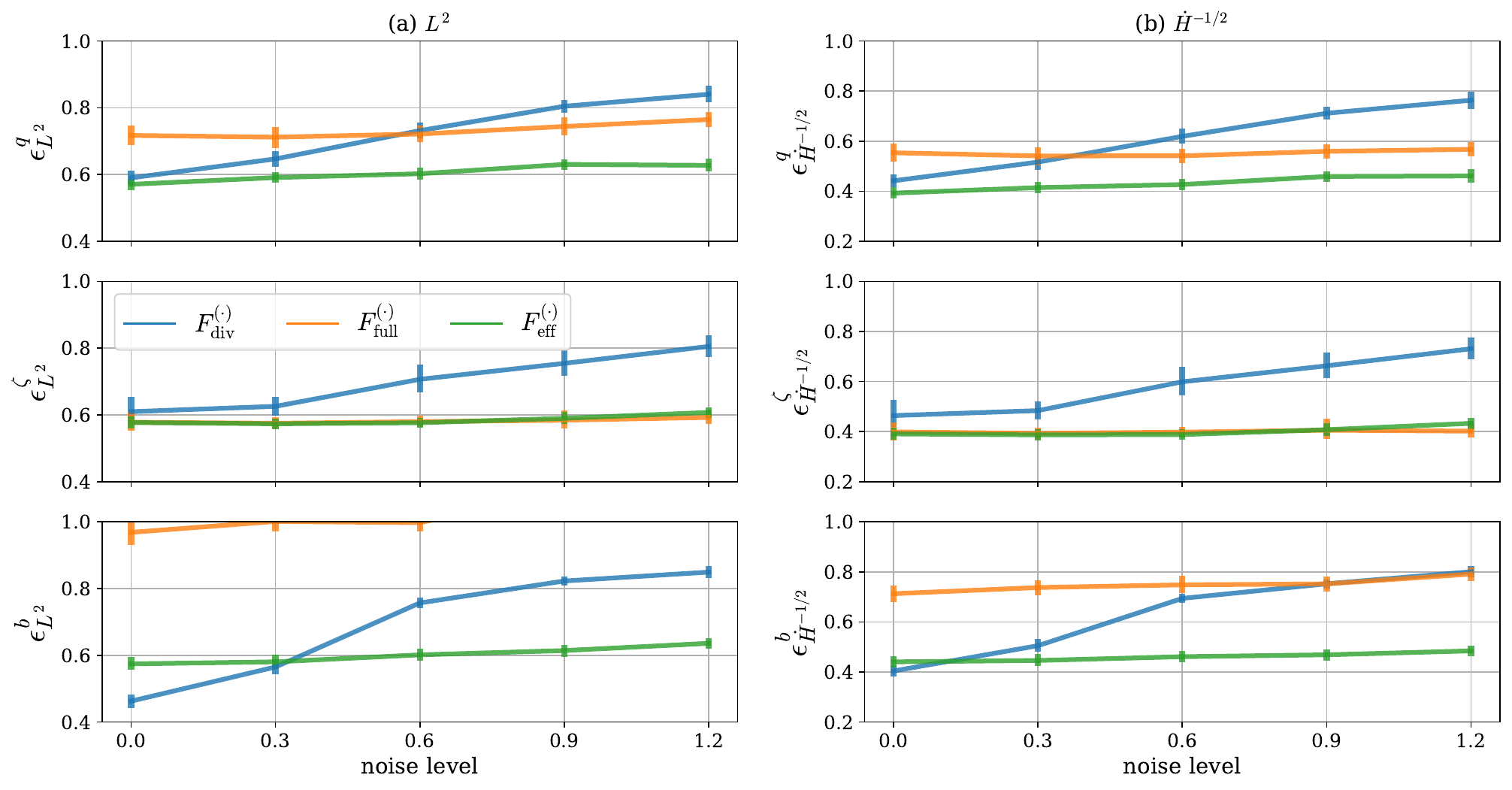}\end{center}
  \caption{Ensemble average and quartiles of the normalized ($a$) $L^2$ norm
  ($b$) $\dot{H}^{-1/2}$ semi-norm, given by Eq.~(\ref{eq:l2_norm}) and
  (\ref{eq:hp_norm}) respectively, for the models predicting the divergence of
  the eddy (rows) PV flux, relative vorticity (cf. momentum) flux, and buoyancy
  flux as a function of noise level (in units of standard deviation of the
  training data), for models using the time-mean streamfunction
  $\overline{\psi}$ as the input. Blue denotes models trained on the divergence
  of the eddy fluxes, orange denotes models trained on the full eddy fluxes, and
  green denotes models trained on the filtered eddy fluxes.}
  \label{fig:f8}
\end{figure}

The $L^2$ and $\dot{H}^{-1/2}$ mismatches of $F^{q/\zeta/b}_{\rm div / full /
eff}(\overline{\psi})$ to the data as a function of noise level for the ensemble
of models is shown in Fig.~\ref{fig:f8}, and consistently we find that the
models trained up on the eddy force function out-perform the models trained upon
the divergence of the eddy flux. The former shows a relative insensitivity to
noise level, while the latter shows a rapid degradation in skill with noise
level. It would seem that the use of eddy force function data alleviates the
sensitivity to small-fluctuations in data, at least in the present measure and
approach.

The reduced sensitivity to noise might have been anticipated, since the eddy
force function is a result of an elliptic solve of a Poisson equation, where the
noisy data is acted upon by an inverse Laplacian operator that leads to
substantial smoothing. We would however argue that the relative insensitivity to
noise is somewhat surprising, since there is no guarantee the presence of even
reduced fluctuations at the streamfunction level would stay small after spatial
derivatives operations, since we are using the divergence of the eddy flux as
the target for the measure of skill. While one could also argue that the present
robustness test is inherently a hard test for models trained upon the divergence
of the eddy flux, we argue the conclusions are robust regardless of whether the
noise is added at the flux, divergence of flux or streamfunction level. In fact,
the use of the divergence of a flux as training data \emph{is} likely the cause
for sensitivity to noise: a inherently small-scale field is sensitive to the
presence of noise in data, so is likely going to lead to issues with robustness.

The conclusions in the above are qualitatively robust for different choices of
the filtering length-scale $L$: with reduced $L$, the degradation of skill in
models trained on the divergence of the eddy fluxes is more rapid with noise
level, but the skill of models trained on the filtered eddy fluxes is still
relatively insensitive to noise level, and consistently more skillful than
models trained on the divergence of the eddy fluxes. The conclusions are also
robust for different choices of inputs ($\overline{\zeta}$ and $\overline{q}$),
and with sample calculations employing other choices of smoothing,
coarse-graining \cite<e.g.,>{Aluie19} or filtering \cite<e.g.,>{Grooms-et-al21}
of the noise field $\eta(x,y)$.

%-------------------------------------------------------------------------------

\section{Conclusions and outlooks}\label{sec:conc}

Data-driven methods are increasingly being employed in problems of Earth System
Modeling, and there is no doubt that such methods provide a powerful tool that
can in principle be leveraged to not only improve our modeling efforts, but also
deepen our underlying understanding of the problems. Most works in the
literature thus far has focused on demonstrating the efficacy of the
machine-learning methods and algorithms. Here we take a complimentary line of
investigation in considering the choice and quality of data itself being fed
into the algorithms, for a case where we have some theoretical understanding to
inform our choice of data. While one could argue this is not entirely necessary
if we just want something that `works' in the relevant metric(s) for the
problem, we argue it is incredibly useful and if not necessary if we want to be
leveraging data-driven methods to learn about the underlying physical problems,
and/or have beyond `black-box' models. Furthermore, the choice of data can in
principle improve the training and/or the performance of the data-driven models
themselves, so there is a need for such an investigation into data quality and
information content.

For this work we focused on the problem of eddy-mean interaction in rotating
stratified turbulence in the presence of boundaries, relevant to the modeling
and parameterization of ocean dynamics. In such systems it is known that the
large-scale mean affects and is affected by the small-scale eddy fluxes, and
while we might want to leverage data-driven methods to learn about the
relationship between the mean and the eddy fluxes, it is known that in the
presence of boundaries the eddy feedback onto the mean is invariant up to a
rotational gauge \cite<e.g.>{MarshallShutts81, FoxKemper-et-al03, Eden-et-al07}.
In practice the dynamically inert component could be quite large \cite<e.g.,>[,
Fig.~\ref{fig:f1} here]{Griesel-et-al09}, and its presence might be expected to
contaminate diagnoses and/or performance of data-driven models. One possible way
round is to train models based on its divergence \cite<e.g.,>{BoltonZanna19,
ZannaBolton20}. Here we propose that data with the dynamically inert eddy fluxes
filtered out could be used instead. The approach outlined here we argued here
may have the advantage in that the resulting field is inherently larger-scale,
which would help with model training and sensitivity, and be theoretically more
appropriate to use if we want to learn about the underlying physics of the
problem, because we do not expect the operations to be commutative (i.e. given
the nonlinearity, learning from the divergence is not guaranteed to be the same
as the divergence of the learned result).

The experimental approach here largely follows that of \citeA{BoltonZanna19},
where we diagnose the various data from a quasi-geostrophic double gyre model to
train the model, and compare the model's performance in its prediction. For
filtering the eddy flux we employ the eddy force function
\cite<e.g.>{MarshallPillar11, Maddison-et-al15, Mak-et-al16b}, which in the
present simply connected quasi-geostrophic system is provably optimal in the
$L^2$ norm \cite<and thus unique; see Appendix of>{Maddison-et-al15}. We made
the choice here to measure a model's skill in being able to reproduce the
divergence of the eddy fluxes over an ensemble of models with 20 members and
over a variety of input choices. The findings here are that the models trained
on the eddy force function are ($a$) more skillful than those trained on the
full eddy flux (except for the relative vorticity eddy fluxes), ($b$) at least
comparable in skill than models trained on the divergence of the eddy fluxes
(except for the buoyancy eddy fluxes), and on occasion better, especially in the
$\dot{H}^{-1/2}$ semi-norm compared to the $L^2$ norm, where the former biases
matching of the large-scale patterns of the resulting predictions, and ($c$)
more robust in that the models are less sensitive to noise in the training data.
The first finding is perhaps not unexpected. The latter two findings we argue
are not entirely obvious, given divergence operations acting at various steps.
For example, sample calculations where a model is trained on the eddy force
function directly (and then taking a Laplacian to obtain a prediction of the
divergence of eddy flux) leads to larger mismatches, which we attribute to the
fact that any mismatches in the predicted eddy force function is significantly
amplified by the two derivative operations. With that in mind, the fact that
models trained on the filtered flux reported here leveraging the eddy force
function as a way to filter data leads to models with comparable or better skill
and superior robustness is a non-trivial result.

Exceptions to the above conclusions are that models trained on the divergence of
the eddy buoyancy flux are more skillful (bottom row of Fig.~\ref{fig:f6}), and
models trained on the eddy relative vorticity flux appear comparable whether the
rotational component is filtered out or not (middle row of Fig.~\ref{fig:f6}).
The former might be justified in that the eddy buoyancy flux is already
relatively smooth and somewhat larger-scale, so that training on its divergence
is not such an issue; however, we also note that the buoyancy eddy force
functions associated with the predictions of model trained on the divergence of
the eddy buoyancy flux seems to perform the worse (bottom of Fig.~\ref{fig:f7}),
implying erroneous predictions of eddy energy pathways. The latter behavior is
possibly to do with the observation that the dynamically inert rotational
component is comparable to the dynamically active divergent component in the
eddy relative vorticity flux (as opposed to the rotational component being a
factor of ten smaller in the eddy PV and buoyancy flux; see
Fig.~\ref{fig:f1}$b,c,e,d$ for eddy PV flux), so the effect of filtering is
somewhat marginal. One saving grace is that, in the quasi-geostrophic system,
the potential vorticity (with contributions from relative vorticity and
buoyancy) is the master variable, and that while models trained up on the
relative vorticity or buoyancy fluxes perform better separately, the models
trained up on the eddy force function has skill and robustness in the master
variable. We note that the conclusions reported here appear to be robust even if
we use data with only some of the rotational component filtered out in sample
calculations (e.g., solving for $\tilde{\psi}$ in Eq.~\ref{eq:helmholtz} with no
normal flux boundary conditions, not shown), although we lose a little bit of
skill and the physical interpretation associated with the eddy force function.

One thing we caution here is drawing a one-to-one comparison of the present work
with that of \citeA{BoltonZanna19} and \citeA{ZannaBolton20}. While it is true
those works utilize a similar model, experimental procedure and data, the main
theoretical difference is that the choice of average is different: their work
utilizes a spatial average, and the eddy flux data there is defined as the
difference between the filtered divergence and the divergence of the filtered
field (if making an assumption of the zero divergence condition on the resulting
velocities). Here we utilize a \emph{time} average, which is in line with the
definition of the eddy force function in \citeA{Maddison-et-al15}, which
requires a Reynolds average. While we have not attempted a similar investigation
in the case of spatial averaging, it is not implausible that there is an
analogous object to the eddy force function when a spatial average is employed,
or that a simple Helmholtz-type decomposition could yield the desired filtering
of the dynamically inert rotational component, but is beyond the scope of the
present investigation.

Because of the choice of time average, we have limited data in time, and one
could wonder whether our conclusions are simply to do with the limited data
availability. This is unlikely the case: we also carried out an analogous
investigation with rolling time averages as well as \emph{ensemble} averages
(not shown), and the conclusions drawn from those results are essentially
identical to those here. This is perhaps not surprising noting that the rolling
time averages for a long enough window and the ensemble averages shown no strong
deviations from each other, but we note this is likely only true for a
sufficiently simple system with no strong evidence of internal modes of
variability, such as the one employed here.

The main intention of the present work is to demonstrate that not all data
choices are equal when fed to data-driven methods, and it is not always
advisable throwing all the available data at the machine and trust that the
machine will figure out what to do with it (although one could argue that might
reduce the inherent biases). For the case of rotating stratified turbulence, the
eddy force function is potentially a useful quantity if we aim to leverage
data-drive methods for model skill or for learning about the underlying physics
of the problem, given the various theoretical expectations highlighted in this
work. Other choices may be possible: in a periodic domain often used in rotating
turbulence studies \cite<e.g.,>{Frezat-et-al22, Ross-et-al23}, a standard
Helmholtz decomposition could be used to solve for the divergent component,
although the eddy force could still be used for physical interpretation. We note
that while skill in reproducing eddy forcing is one target, we have not examined
here on the ability of the model to reproduce the mean state, and the present
procedure might be termed an `offline' approach. Learning `online'
\cite<e.g.,>{Frezat-et-al22} may be more appropriate for parameterization
purposes to improve on the mean response, and it would be of interest to see
whether filtering of the eddy flux as discussed here would confer any benefits
to model learning.

The present work also highlights questions relating to information content of
data. While quantifying absolute data information content is likely quite
difficult, it should be at least possible to compute a relative measure, even if
empirically. Preliminary investigation indicates that as the amount of data
exposed to the machine learning algorithm is reduced, the accuracy of models
trained upon the full eddy flux or the divergence of the eddy flux degrades much
faster than models trained upon the eddy force function. One might ask an
analogous question of the input data. The work of \citeA{BoltonZanna19} suggests
for example that training with data from regions with higher eddy kinetic energy
leads to better model performance in terms of accuracy, suggestive of higher
information content in said region. Within the present experimental framework,
instead of training using all the data and performing a random sampling of the
sub-regions considered in this work, we could consider instead not using all the
data, and perform training based on a biased sampling that favor regions with
higher eddy energy content, with the hypothesis that the latter case leads to
models with higher accuracy from a statistical point of view. Further, we could
investigate the case of multiple inputs, where we hypothesize that eddy energy
and a mean state variable as inputs might lead to improved performance compared
to say two mean state variables: in the current quasi-geostrophic setting, the
mean state variables are functionally related to each other, possibly leading to
redundant information, while the eddy energy might be dependent on the mean
state, but is capturing eddy statistics instead and providing complementary
information. This investigation is ongoing and will be reported elsewhere in due
course.

%\begin{figure}
%  \includegraphics[width=\textwidth]{figures/Fig7}
%  \caption{robustness}
%  \label{fig:f7}
%\end{figure}

%-------------------------------------------------------------------------------

\section*{Data Availability Statement}

This work utilizes FEniCS (\texttt{2019.1.0}) that is available as a Python
package. The source code for the model (\verb|qgm2|, from James Maddison),
sample model data and scripts used for generating the plots in this article from
the processed data are available through
\url{http://dx.doi.org/10.5281/zenodo.8072817}.

\acknowledgments

This research was funded by both RGC General Research Fund 16304021 and the
Center for Ocean Research in Hong Kong and Macau, a joint research center
between the Qingdao National Laboratory for Marine Science and Technology and
Hong Kong University of Science and Technology. We thank James Maddison and
Liiyung Yeow for various scientific and technical comments in relation to the
present investigation, and the former for providing the \verb|qgm2| code for use
in the present work.

%% ------------------------------------------------------------------------ %%
%% References and Citations

%%%%%%%%%%%%%%%%%%%%%%%%%%%%%%%%%%%%%%%%%%%%%%%
%
% \bibliography{<name of your .bib file>} don't specify the file extension
%
% don't specify bibliographystyle
%%%%%%%%%%%%%%%%%%%%%%%%%%%%%%%%%%%%%%%%%%%%%%%

\bibliography{refs}

%Reference citation instructions and examples:
%
% Please use ONLY \cite and \citeA for reference citations.
% \cite for parenthetical references
% ...as shown in recent studies (Simpson et al., 2019)
% \citeA for in-text citations
% ...Simpson et al. (2019) have shown...
%
%
%...as shown by \citeA{jskilby}.
%...as shown by \citeA{lewin76}, \citeA{carson86}, \citeA{bartoldy02}, and \citeA{rinaldi03}.
%...has been shown \cite{jskilbye}.
%...has been shown \cite{lewin76,carson86,bartoldy02,rinaldi03}.
%... \cite <i.e.>[]{lewin76,carson86,bartoldy02,rinaldi03}.
%...has been shown by \cite <e.g.,>[and others]{lewin76}.
%
% apacite uses < > for prenotes and [ ] for postnotes
% DO NOT use other cite commands (e.g., \citet, \citep, \citeyear, \nocite, \citealp, etc.).
%

\end{document}